\documentclass[aps,prb,showpacs,groupedaddress,superscriptaddress,twocolumn]{revtex4}

\usepackage{graphicx,amsmath}
\usepackage{color}
\usepackage{float}
\begin{document}

\preprint{APS/123-QED}

\title{Global Phase Diagram of Disordered Type-II Weyl Semimetals}

\author{Yijia Wu}
\affiliation{International Center for Quantum Materials, School of Physics, Peking University, Beijing 100871, China}

\author{Haiwen Liu}
\affiliation{Center for Advanced Quantum Studies, Department of Physics, Beijing Normal University, Beijing 100875, China }

\author{Hua Jiang}
\thanks{\texttt{jianghuaphy@suda.edu.cn}}
\affiliation{College of Physics, Optoelectronics and Energy, Soochow University, Suzhou 215006, China}

\author{X. C. Xie}
\affiliation{International Center for Quantum Materials, School of Physics, Peking University, Beijing 100871, China}
\affiliation{Collaborative Innovation Center of Quantum Matter, Beijing 100871, China}

\date{\today}

\begin{abstract}
With electron and hole pockets touching at the Weyl node, type-II Weyl semimetal is a newly proposed topological state distinct from its type-I cousin. We numerically study the localization effect for tilted type-I as well as type-II Weyl semimetals and give the global phase diagram. For disordered type-I Weyl semimetal, an intermediate three-dimensional quantum anomalous Hall phase is confirmed between Weyl semimetal phase and diffusive metal phase. However, this intermediate phase is absent for disordered type-II Weyl semimetal. Besides, near the Weyl nodes, comparing to its type-I cousin, type-II Weyl semimetal possesses even larger ratio between the transport lifetime along the direction of tilt and the quantum lifetime. Near the phase boundary between the type-I and the type-II Weyl semimetals, infinitesimal disorder will induce an insulating phase so that in this region, the concept of Weyl semimetal is meaningless for real materials.
\end{abstract}

\maketitle



\section{Introduction}

The Weyl semimetal (WSM) is a class of novel quantum state with Weyl cones appear in pairs in its bulk spectrum and Fermi arcs on the surface. WSM can be divided into two types: type-I Weyl semimetal (WSM1) with vanishing density of states (DOS) at the Weyl node and type-II Weyl semimetal (WSM2) with finite DOS contributed by electron and hole pockets separated by the Weyl node. The earlier proposals for candidate materials of WSM are all WSM1 \cite{XianGang_Wan, Ran_Ying_PRB, HongMing_Weng_PhysRevX}. In case that Lorentz invariance is broken, the Weyl cones may be tipped over and transformed into WSM2. The first theoretical prediction for possible WSM2 materials is WTe$_2$ \cite{XiDai_typeII_theory}. Some experimental evidences for WTe$_2$ as WSM2 have been observed \cite{WTe2_exp_1, WTe2_exp_2, WTe2_exp_3, WTe2_exp_4}. MoTe$_2$ is another theoretically predicted candidate of WSM2 \cite{MoTe2_1, MoTe2_2} and experimental confirmations have been reported \cite{Shuyun_Zhou_typeII_exp, MoTe2_exp_ARPES}. The Landau level (LL) structure of type-II Weyl cone depends on the direction of the magnetic field \cite{XiDai_typeII_theory, Hybrid_WSM}. With anisotropic LL structure and coexistence of electron and hole pockets at the Fermi energy, WSM2 exhibits many interesting phenomena in the presence of a magnetic field \cite{Magnetic_WSMII_1, Magnetic_WSMII_2, Magnetic_WSMII_3}.

In the absence of a magnetic field, the disorder-induced phase transition is widely observed for a large number of quantum systems, such as the well-known metal-insultor transition in dirty metals \cite{Gang_of_four, TMM_1} and the topological Anderson insulator phase in several disordered systems  \cite{Topological_Anderson_insulator, Topological_Anderson_insulator_HuaJiang, Topological_Anderson_insulator_3D}. However, comparing to WSM1 (without the tilt term) whose global phase diagram have been thoroughly investigated \cite{Chui-Zhen_PRL, Shindou_disorderedWSM, Hughes_disorderedWSM, WSM1_Sau_Roy}, disorder-induced phase transitions for tilted WSM, especially WSM2, have not been paid much attention yet. The diffusive phase has been reported in the presence of disorder for single tilted type-I Weyl cone \cite{k_dependent_SCBA}. Nevertheless, the ``no-go theorem" guarantees that the Weyl nodes always appear pairwise \cite{no_go_1, no_go_2}. Phase transition from WSM1 to WSM2 is also confirmed since disorder may renormalize the topological mass \cite{Moon_Jip_Parl}. However, the global phase diagram of tilted Weyl semimetals in the presence of disorder still needs to be explored.

In this work, we investigate the disorder-induced multiple phase transitions for both tilted WSM1 and WSM2 in the presence of an inversion-like symmetry. Transfer matrix method is used for numerically calculating the normalized localization length. Self-consistent Born approximation (SCBA) is adopted to understand the phase transitions by an analytical approximation. The results of the transfer matrix method give the global phase diagram, which shows that plenty of quantum states may exist in the disordered WSM systems. By combining both the numerical and analytical results, it is shown that along the direction of tilt, WSM2 is more robust against disorder than WSM1, furthermore, it possesses larger lifetime ratio (defined as the ratio between transport lifetime and quantum lifetime) near the Weyl nodes.

This paper is organized as follows. In Sec. II, we introduce the model Hamiltonian and the corresponding symmetries. A clean phase diagram is also shown here. In Sec. III, we present transfer matrix method to show the localization effect and global phase diagram in the presence of disorder, and in Sec. IV, we use SCBA to show the disorder-induced renormalization and band broadening. Finally, a brief summary, as well as some discussions and proposals are given in Sec. V.


\section{Model and clean phase diagram}

The tilted WSM can be described by a simple $2\times2$ Hamiltonian $H=H_{0}+H_{tilt}$,
where $H_0$ describes type-I Weyl semimetal preserving particle-hole symmetry \cite{Ran_Ying_PRB, Chui-Zhen_PRL}

\begin{equation}
\begin{split}
\label{H0}
H_{0} =& \left(m_{z}-t_{z}\cos k_{z}\right)\sigma_{z}+m_{0}\left(2-\cos k_{x}-\cos k_{y}\right)\sigma_{z} \\
&+ t_{x}\sigma_{x}\sin k_{x}+t_{y}\sigma_{y}\sin k_{y}
\end{split}
\end{equation}

\noindent and $H_{tilt}$ introduces two tilt terms \cite{Moon_Jip_Parl}

\begin{equation}
\label{Htilt}
H_{tilt}=a_t\sin(k_z)I_{2 \times 2}+b_t\cos(k_z)I_{2 \times 2}
\end{equation}

\noindent where both $a_{t}$ and $b_{t}$ are real numbers and $I_{2\times2}$ denotes 2 by 2 identity matrix.

Both time-reversal (TR) symmetry and inversion symmetry are broken in $H_0$ \cite{Hybrid_WSM}: $\mathcal{T}H_{0}\left(\mathbf{k}\right)\mathcal{T}^{-1}\neq H_{0}\left(-\mathbf{k}\right)$ and $\mathcal{P}H_{0}\left(\mathbf{k}\right)\mathcal{P}^{-1}\neq H_{0}\left(-\mathbf{k}\right)$,
where the TR and inversion operators are defined as $\mathcal{T}=\sigma_{y}K$ and $\mathcal{P}=I_{2\times2}$, respectively ($K$ denotes the complex conjugation operator). However, it is easy to check that $H_{0}$ preserves emergent inversion-like symmetry and antiunitary particle-hole symmetry \cite{Hybrid_WSM}, {\it i.e.} $\sigma_{z}H_{0}\left(\mathbf{k}\right)\sigma_{z}=H_{0}\left(-\mathbf{k}\right)$ and $\sigma_{x}H_{0}\left(\mathbf{k}\right)\sigma_{x}=-H_{0}^{*}\left(-\mathbf{k}\right)$, respectively. In the presence of these two symmetries, two type-I Weyl nodes appear at energy $E=0$ in the absence of $H_{tilt}$.

\begin{figure}[b]
\includegraphics[width=0.33\textwidth]{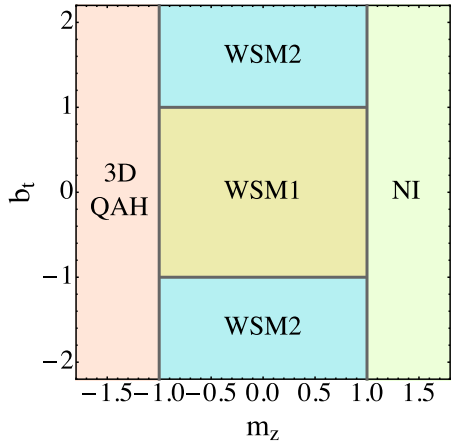}
\caption{\label{clean_phase_diagram_WSM} (color online). Phase diagram in the clean limit $W=0$. Other parameters are $t_x=t_y=t_z=1$, $m_0=2.1$ and $a_t=0$.}
\end{figure}

Considering the Hamiltonian of tilted WSM $H=H_0+H_{tilt}$, here non-zero $a_{t}$ breaks the emergent inversion-like
symmetry $\sigma_{z}H\left(\mathbf{k}\right)\sigma_{z}=H\left(-\mathbf{k}\right)$ and non-zero $b_{t}$ breaks the antiunitary particle-hole symmetry $\sigma_{x}H\left(\mathbf{k}\right)\sigma_{x}=-H^{*}\left(-\mathbf{k}\right)$
. Hence, in the case of $a_{t}\neq0$ and $b_{t}\neq0$, one may get hybrid Weyl semimetal \cite{Hybrid_WSM} that one Weyl cone is type-I while the other is type-II and these two Weyl nodes appear at different energies. Throughout this work, we focus on the condition that $a_{t}=0$. In this case, emergent inversion-like symmetry gives us $E_{\pm}(\mathbf{k}) = E_{\pm}(-\mathbf{k})$ (here $E_{\pm}$ denotes the energies of the upper and lower bands), so that two Weyl cones are of the same type and these two Weyl nodes appear at the same energy. In this way, one can avoid the complication of hybrid WSM, and clearly show the differences between WSM1 and WSM2 in the presence of disorder. Besides, one can locate the Fermi energy at the Weyl nodes and investigate the intervalley scattering without unwanted interference from the contribution of the trivial states.

In addition, the disorder effect is introduced by adding the random on-site potential $H_{dis}$ to $H$, where

\begin{equation}
\label{Hdis}
H_{dis} =
    \begin{pmatrix}
    V_1(\mathbf{r}) & 0 \\
    0 & V_2(\mathbf{r})
    \end{pmatrix}\\
\end{equation}

\noindent both $V_{1,2}(\mathbf{r})$ are uniformly distributed in the range of $[-W/2, W/2]$ with $W$ represents the strength of disorder.

\begin{figure}[b]
\includegraphics[width=0.48\textwidth]{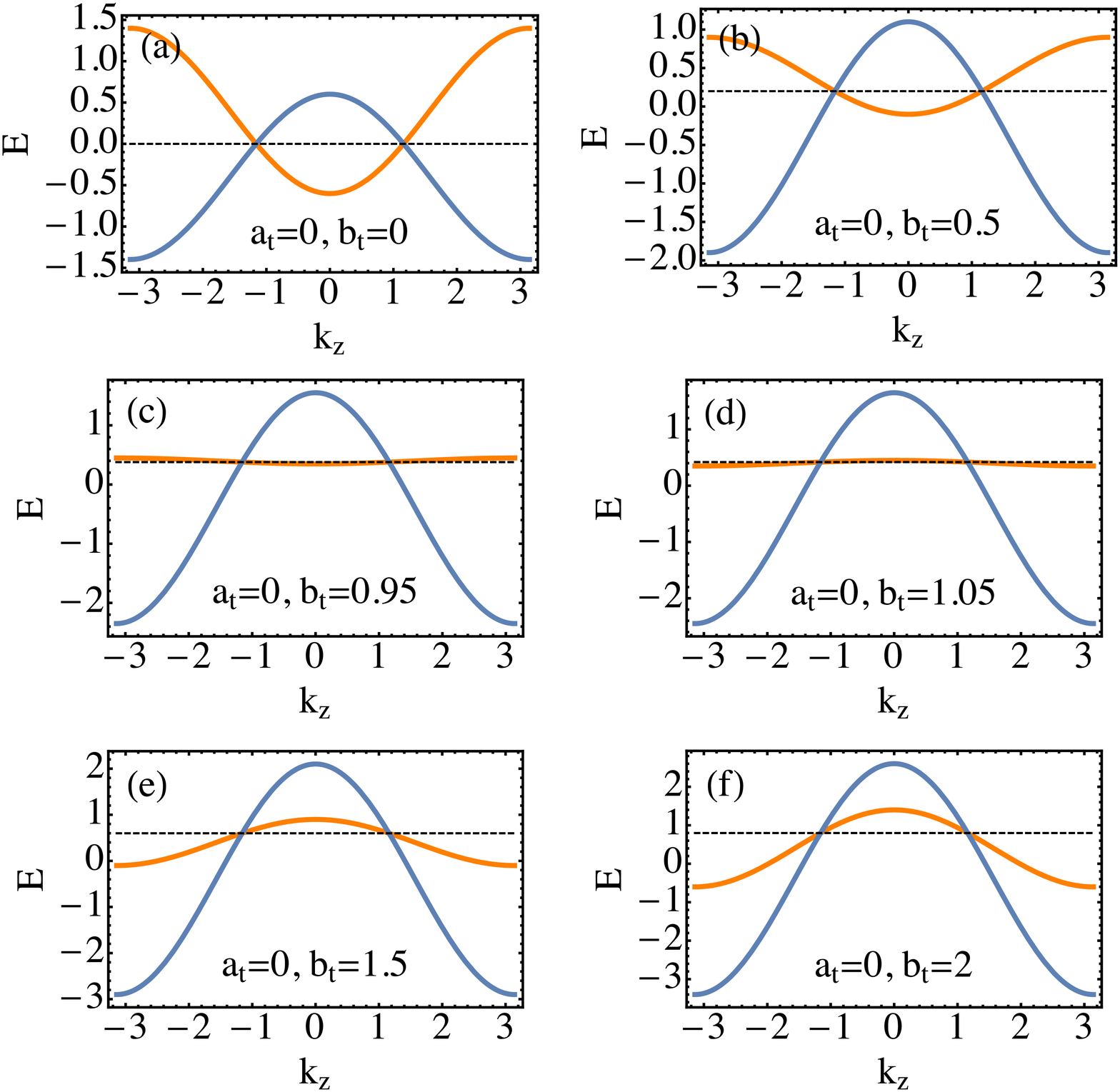}
\caption{\label{dispersion} (color online). (a)-(f) Band structures $E \sim k_z$ ($k_z \in [-\pi, \pi]$) in the plane of $k_x=k_y=0$ for $a_t=0$ and different $b_t$'s. Dashed lines denote the Fermi levels. Other parameters are $t_x=t_y=t_z=1$, $m_0=2.1$ and $m_z/m_0=0.190$. The phase transition point between WSM1 and WSM2 is $b_t=1$.}
\end{figure}

\begin{figure*}[!t]
\centering
\includegraphics[width=0.96\textwidth]{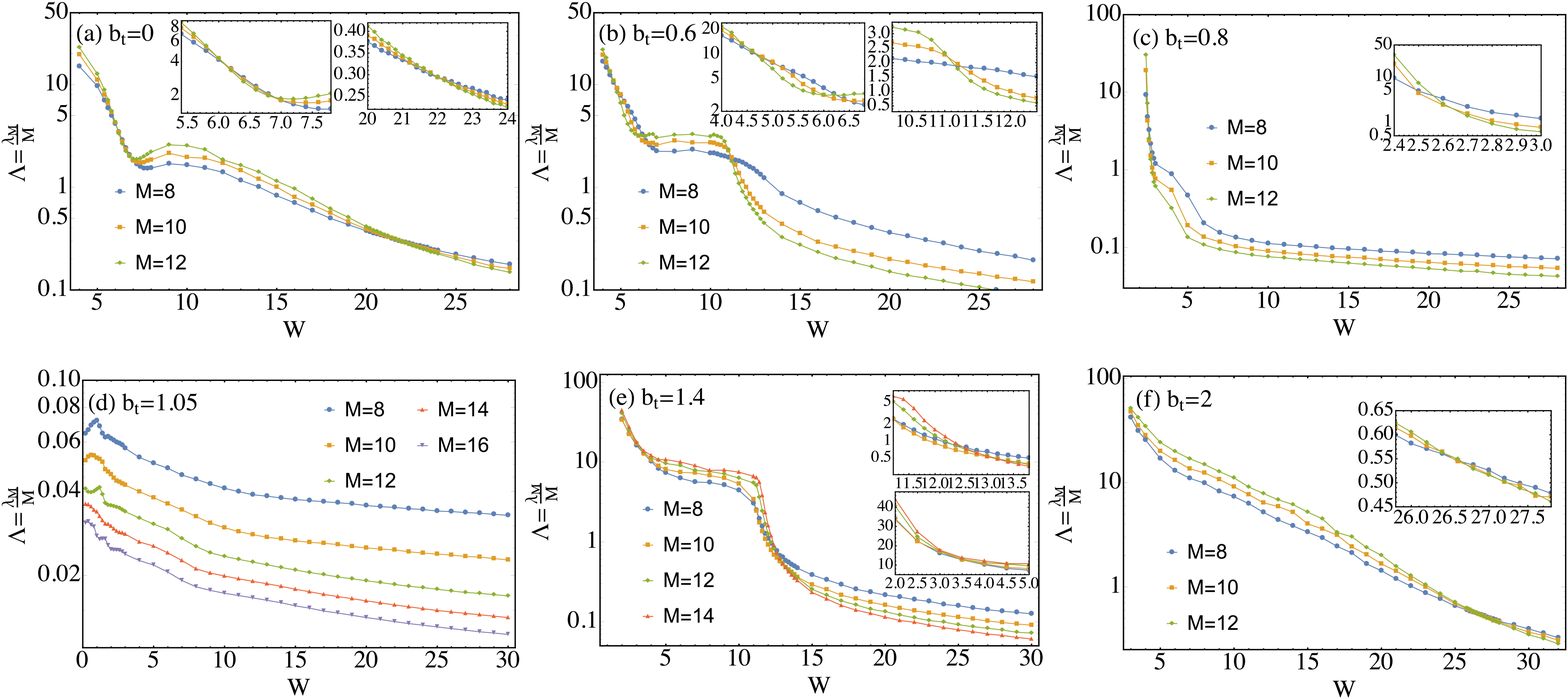}
\caption{\label{TMM_result} (color online). (a)-(f) Results of the transfer matrix method for normalized localization length $\Lambda$ versus disorder strength $W$ at different tilt strength $b_t$ and different sample width $M$. The insets show details near the phase transition points. Other parameters are $t_x=t_y=t_z=1$, $m_0=2.1$, $m_z/m_0=0.190$ and $a_t=0$.}
\end{figure*}

\begin{figure}[!b]
\includegraphics[width=0.43\textwidth]{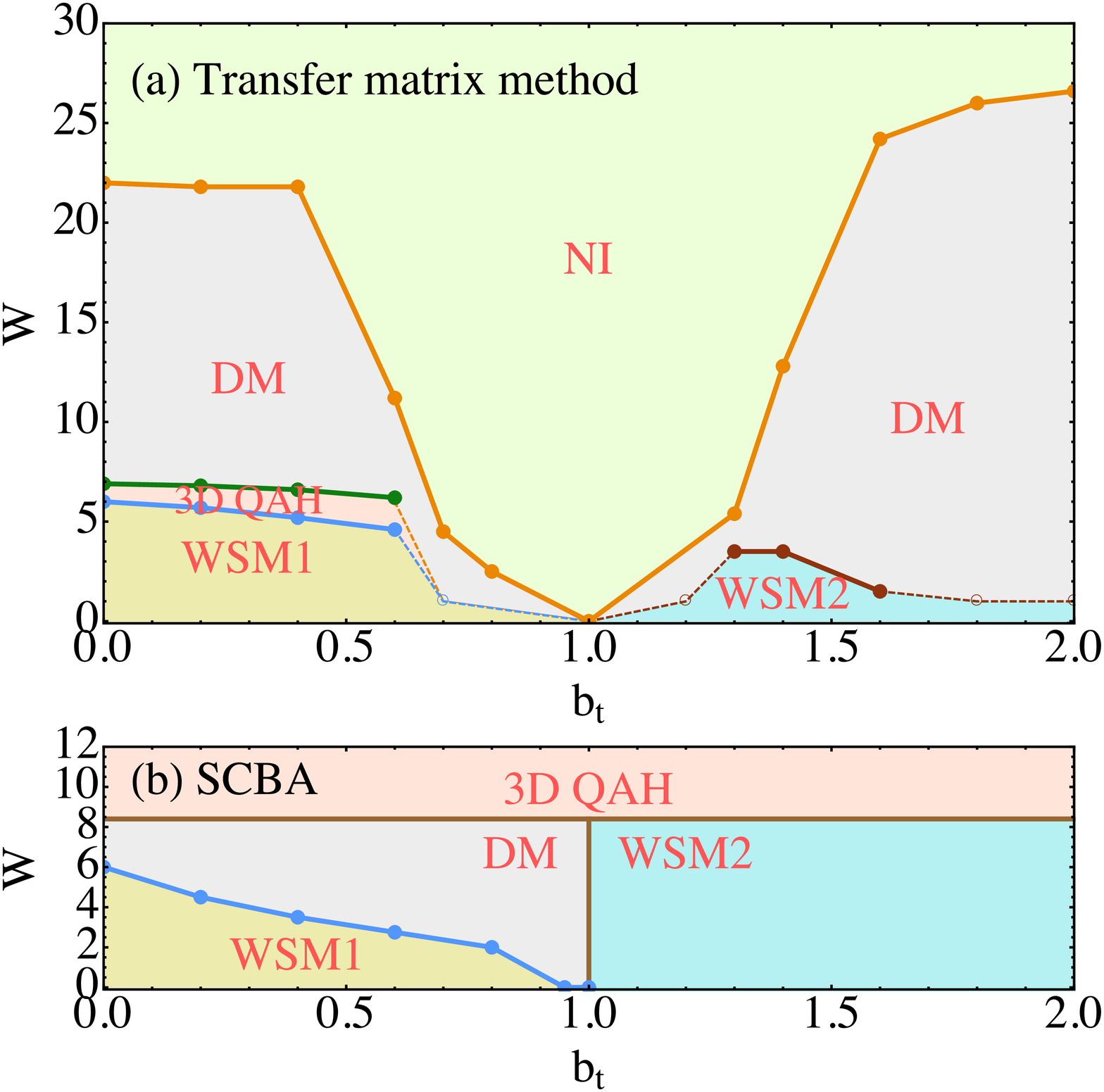}
\caption{\label{phase_diagram} (color online). Phase diagram on the $W \sim b_t$ plane for $m_z/m_0=0.190$, $m_0=2.1$, $t_x=t_y=t_z=1$ and $a_t=0$. (a) Phase diagram determined by the transfer matrix method. The filled circles are phase transition points identified by the transfer matrix method. The solid lines are the phase boundaries connected by filled circles. Empty circles and dashed lines represents suspected phase boundaries in the regions where the localization lengths are too long to get precise results. The intermediate 3D QAH phase is regarded as the natural extension of the previously determined 3D QAH phase \cite{Chui-Zhen_PRL}. (b) Phae diagram determined by SCBA. The filled circles indicate the critical disorder strength $W_{c1}$ separating WSM1 phase and DM phase.}
\end{figure}

In the following, we choose $t_x=t_y=t_z=1$ and $m_0=2.1$ \cite{Chui-Zhen_PRL}. The phase diagram in the clean limit $W=0$ is shown in Fig. \ref{clean_phase_diagram_WSM}. In the case of $m_z<-t_z$, two Weyl nodes are annihilated at the Brillouin zone (BZ) boundary. Two bands are gapped and the band inversion gives us the three-dimensional quantum anomalous Hall (3D QAH) state \cite{Chui-Zhen_PRL}. On the contrary, in the case of $m_z>t_z$, two Weyl nodes are annihilated pairwise at the center of the BZ and then we have a normal insulator (NI) phase. Finally, within $-t_z \leq m_z \leq t_z$, the system is gapless so one gets WSM phase with two Weyl nodes. Considering the effective Fermi velocity $\bar{v}$ near the Weyl nodes \cite{XiDai_typeII_theory, Hybrid_WSM}, WSM1 phase appears in the condition for $-1 < b_t < 1$ and WSM2 appears in the condition for $b_t < -1$ or $b_t > 1$ (Notcing that $a_t=0$ is fixed as mentioned before).

 To have an intuitive illustration, the band structures $E \sim k_z$ for $m_z/m_0 = 0.190$ and $b_t=0, 0.5, 0.95, 1.05, 1.5$ and $2$ in the plane of $k_x=k_y=0$ are plotted in Fig.\ref{dispersion}. Emergent inversion-like symmetry guarantees that two Weyl cones are of the same type and appear at the same energy. Two Weyl cones are gradually tilted with increasing $b_t$. $b_t=1$ is the phase transition point between WSM1 and WSM2 in the clean limit.


\section{Numerical Results by The Transfer Matrix Method}

The transfer matrix method \cite{TMM_1, TMM_2} is widely adopted for numerical calculations of the localization lengths and determination of the phae boundaries for disordered condensed matter systems. A great advantage of the transfer matrix method is that no approximation is presented in its formalism. The only error comes from the finite size effect.

We numerically calculate the localization length along the direction of tilt ($z$-direction in our work) by using the transfer matrix method. 3D long bar samples with square cross-section $M_x=M_y=M$ and of length $M_z$ are considered. Periodic boundary conditions are adopted both in the $x$ and $y$ directions. The normalized localization lengths $\Lambda = \lambda_M / M$ are investigated for different sample width $M$. Typically, $\Lambda$ increases with $M$ for metallic phase and decreases with $M$ for insulating phase. The Fermi energy $E_F$ in the transfer matrix method calculation is chosen at the energy of the Weyl nodes, and the relative errors \cite{TMM_relative_error, FrankMilde} for all the $\Lambda$'s in this work are less than $3\%$.

Typical results for normalized localization length $\Lambda$ versus disorder strength $W$ at different tilt strengths $b_t$ are shown as Fig. \ref{TMM_result}. The global phase diagram determined by transfer matrix method is summarized in Fig. \ref{phase_diagram}(a), which is the main result of our paper. There are several kinds of multiple phase transitions for different $b_t$'s with increasing of disorder strength $W$.

For small $b_t $ such as $0, 0.2, 0.4$ and $0.6$, we have two type-I Weyl cones in the clean limit. With increasing of disorder strength $W$, the system will go through 3D QAH phase, diffusive metal (DM) phase and then NI phase in sequence. Remarkably, the intermediate 3D QAH phase is not quite apparent at $b_t=0$ (see Fig. \ref{TMM_result} (a)) and it was missed in the previous global phase diagram of WSM1 \cite{Chui-Zhen_PRL}, whereas it has been recently reported in a detailed study of disordered WSM1 with particle-hole symmetry ($a_t=b_t=0$) \cite{XiangRong_Wang}. Furthermore, this intermediate 3D QAH phase becomes more apparent with a finite tilt strength $b_t$, since $|d\Lambda / dM|$ gets larger in this 3D QAH phase with increasing $b_t$ (e.g. comparing Fig. \ref{TMM_result} (a), (b)). This phase can be understood as the pairwise annihilation of Weyl nodes by the disorder-induced scattering between two Weyl nodes. It can be inferred that this intermediate phase is a natural extension of the previously determined 3D QAH phase and the phase boundary between 3D QAH phase and WSM1 phase needs to be modified (see Fig. 2 of Ref. \cite{Chui-Zhen_PRL}).

For larger tilt strength such as $b_t=0.7$ and $0.8$, only the phase transition between DM phase and NI phase (Anderson transition) is observed through the transfer matrix method (see Fig. \ref{TMM_result}(c)). This is surprising since in the clean limit $W=0$, there ought to be a WSM1 phase. Furthermore, 3D QAH phase may still exists between the DM phase and WSM1 phase. For small disorder strength $W$, however, the normalized localization lengths $\Lambda$ are too long (typically, $\sim 10^2$) to get accurate data through the transfer matrix method. Thus, we use empty circles and dashed lines to represent expected phase boundaries in this region (see Fig. \ref{phase_diagram}(a)). In addition, in the range of $0<b_t<1$, the critical disorder strength for Anderson transition becomes smaller with increasing of the tilt strength $b_t$.

Interestingly, when $b_t$ approaches $1$ (no matter from the WSM1 side or from the WSM2 side), the metallic phase is absent for all disorder strengths $W$ investigated (see Fig. \ref{TMM_result}(d)). Besides, the normalized localization lengths are extremely short. An interesting consequence of these results could be that an infinitesimal disorder will induce insulating phase when $b_t$ is approximate to $1$. It can be explained that one band is nearly flat here (see Fig.\ref{dispersion} (c), (d)). The flat band is insulating since the band mass of electron is extremely large. Noticing that ``big-$k$" scattering is allowed in our model since the disorder potential is on-site. Thus the electron in another non-flat band is also localized due to the band coupling through the disorder-induced intervalley scattering. Therefore, in real materials, the concept of Weyl semimetal may be meaningless near the phase boundary between WSM1 and WSM2 since the WSM phase will be destroyed by an infinitesimal disorder.

At $b_t=1.3, 1.4$ and $1.6$, besides the Anderson transition, the results of the transfer matrix method also show that two metallic phases are separated by a phase transition point where $d\Lambda / dM = 0$ (see Fig. \ref{TMM_result}(e)). Since we ought to get WSM2 in the clean limit $W=0$, it can be reasonably suggested that these two phases are WSM2 phase in the weak disorder region and DM phase in the strong disorder region, respectively. However, for small $b_t$ such as $b_{t}=1.2$ or large $b_t$ such as $1.8, 2$, the normalized localization length again gets very large for weak disorder strength $W$, so one cannot ontain realiable results by adopting the transfer matrix method in these regions. Nevertheless, no evidence show the existence of intermediate 3D QAH phase between the DM phase and WSM2 phase. In WSM1, the emergency of this intermediate 3D QAH phase is explained as the annihilation of two Weyl nodes by the disorder scattering and the opening of a topologically nontrivial gap. However, in case of WSM2 with non-zero DOS at the Fermi level, the scattering between two Weyl nodes are no longer as important as in WSM1. This intermediate 3D QAH phase is expected to be absent in WSM2. In another word, we cannot get an insulating phase by relatively weak disorder since there is a non-zero DOS for WSM2 in the clean limit. Besides, the normalized localization length is typically longer in WSM2 than in WSM1 (for example, comparing the normalized localization length for $b_t=0$ and $b_t=2$ at the same width $M$ and same disorder strength $W$, see Fig. \ref{TMM_result} (a), (f)). We can conclude that WSM2 is more robust against disorder and it might be more useful in terms of practical applications.


\section{Self-consistent Born approximation}

The SCBA \cite{Beenakker_SCBA} is widely used in dealing with disorderd WSM systems \cite{Moon_Jip_Parl, Mikito_k_dependent_SCBA, k_dependent_SCBA, Chui-Zhen_PRL}. The self-energy obtained by SCBA can be considered as the disorder-induced modification for the Hamiltonian. The real part of self-energy renormalizes the parameters of the original Hamiltonian, while the imaginary part determines the band broadening and decaying of the quasi-particles.

The self-energy $\Sigma(\mathbf{k}, E_F)$ in the scheme of SCBA reads

\begin{equation}
\label{self_energy}
\Sigma(\mathbf{k}, E_F) = \int_{\mathbf{k'} \in BZ} \frac{d^3 \mathbf{k'}}{(2\pi)^3} n_i |u(\mathbf{k}-\mathbf{k'})|^2 \times \langle G(\mathbf{k'}, E_F) \rangle
\end{equation}

\noindent where $n_i=1$ is the number of scatterers per unit volume and the averaged Green's function $\langle G(\mathbf{k}, E_F) \rangle$ can be expressed as $ \langle G(\mathbf{k'}, E_F) \rangle = [ (E_F+i0^+)I_{2 \times 2}- H(\mathbf{k'})- \Sigma(\mathbf{k'}, E_F) ]^{-1} $. The disorder potential in our model is on-site and evenly distributed in the ragne of $[-W/2, W/2]$, so that the disorder potential in $k$-space

\begin{multline}
\label{u_of_k}
|u(\mathbf{k}-\mathbf{k'})|^2 = \frac{1}{W} \int_{-W/2}^{W/2} dw \cdot \left| \int d^3 \mathbf{r} \cdot e^{i(\mathbf{k}-\mathbf{k'}) \cdot \mathbf{r}} w \delta( \mathbf{r} ) \right|^2 \\
= \frac{1}{W} \int_{-W/2}^{W/2} dw \cdot w^2 = \frac{W^2}{12}
\end{multline}

\noindent is actually $k$-independent. Consequently, the self-energy $\Sigma(\mathbf{k}, E_F)$ is also $k$-independent and satisfies the self-consistent equation as follow

\begin{multline}
\label{k_independent_SCBA}
\Sigma (E_F) = \frac{W^2}{96 \pi^3} \times \\ \int_{\mathbf{k'} \in BZ} d^3 \mathbf{k'} \frac{1}{ (E_F+i 0^+)I_{2 \times 2}- H(\mathbf{k'}) -\Sigma(E_F) }
\end{multline}

Noticing that the self-energy is a $2 \times 2$ matrix and can be decomposed into $\Sigma(E_F) = \Sigma_0 I_{2 \times 2} + \Sigma_x \sigma_x + \Sigma_y \sigma_y + \Sigma_z \sigma_z$. However, both $\Sigma_x$ and $\Sigma_y$ vanish since both $\sigma_x$ term and $\sigma_y$ term are odd in the Hamiltonian $H$. Therefore, in the following, we only need to deal with the real and imaginary parts of $\Sigma_0$ and $\Sigma_z$.

\begin{figure}
\includegraphics[width=0.49\textwidth]{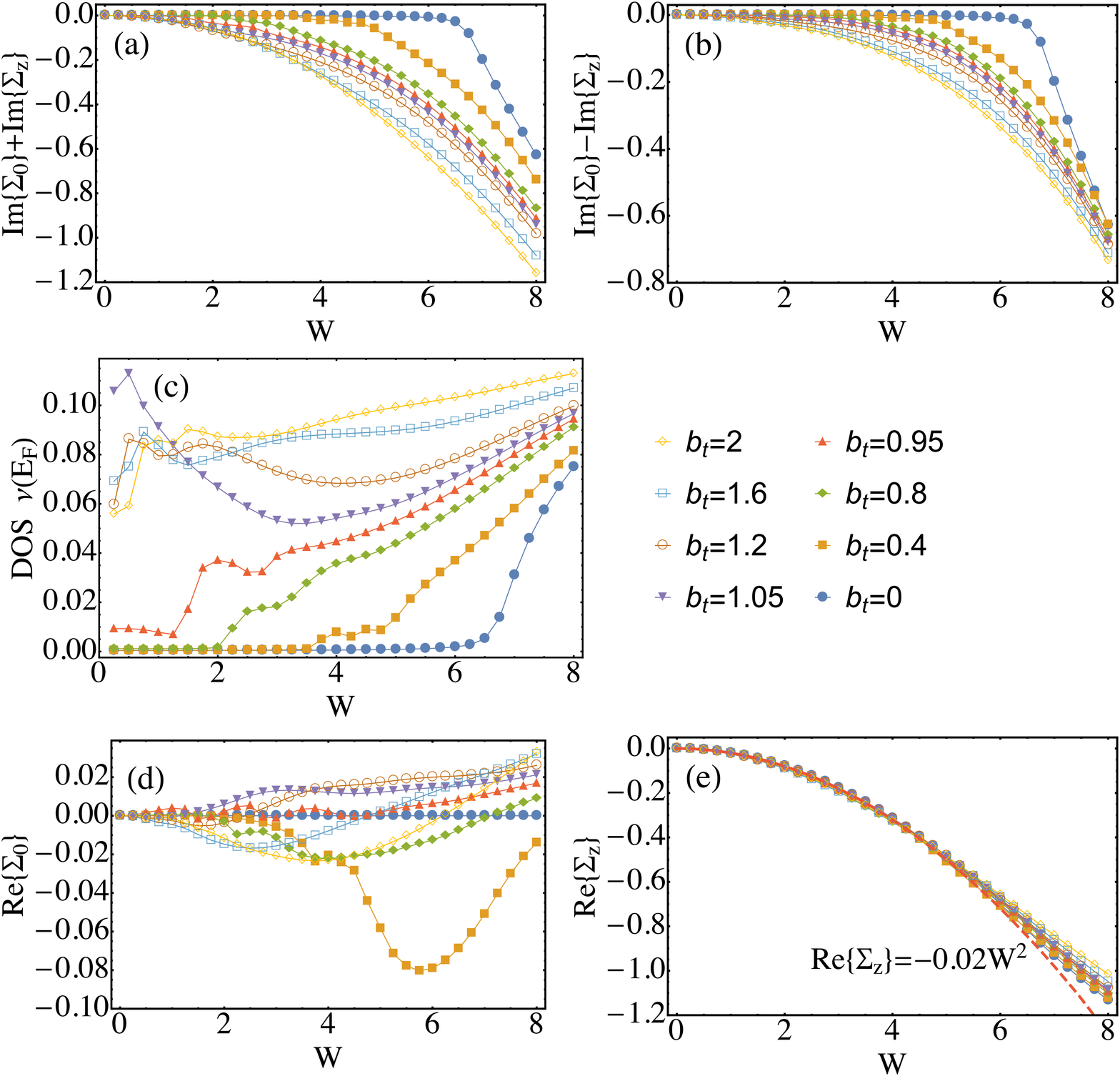}
\caption{\label{SCBA} (color online). (a)-(b) The imaginary part of the diagonal elements of the self-energy $Im{ \{\Sigma_0\} } \pm Im{ \{\Sigma_z\} }$ versus disorder strength $W$. (c) The DOS at the Fermi level $\nu(E_F)$ versus disorder strength $W$. (d)-(e) The real part of the self-energy $Re{ \{\Sigma_0\} }$ and $Re{ \{\Sigma_z\} }$ versus disorder strength $W$. $Re{ \{\Sigma_0\} }$ shifts the Fermi level. $Re{ \{\Sigma_z\} }$ renormalizes the topological mass $m_z$, which can be fitted by $Re{\{\Sigma_z\}} = -0.02 W^2$.}
\end{figure}

The inverse of $Im{\{\Sigma(E_F)\}}$ determines the quantum lifetime of the quasi-particles. Both $Im{ \{ \Sigma_0 \} } \pm Im{ \{ \Sigma_z \} }$ are negative and decrease with increasing of the tilt term $b_t$ and the disorder strength $W$ (See Fig. \ref{SCBA} (a), (b)). It indicates that there are two kinds of quasi-particles in the disordered WSM, their quantum lifetimes both decrease with increasing of $b_t$ and $W$. Besides, the DOS at Fermi level $\nu(E_F)$ is related to the imaginary part of $\Sigma_0$ as

\begin{multline}
\label{DOS}
\nu(E_F) = -\frac{1}{\pi} Im{\left\{ \int \frac{d^3\mathbf{k}}{(2\pi)^3} Tr[\langle G(\mathbf{k'}, E_F) \rangle] \right\}} \\
= -\frac{12}{\pi W^2} Im{\{ Tr[\Sigma (E_F)] \}} = -\frac{24}{\pi W^2} Im{\{ \Sigma_0 \}}
\end{multline}

In the case of $b_t<1$, the DOS at the Fermi level vanishes in the clean limit since both Weyl cones are type-I. A critical disorder strength $W_{c1}$ is observed in the results of SCBA (See Fig. \ref{SCBA} (c)). In the case of $W < W_{c1}$, one still has WSM1 with vanishing DOS at the Fermi level. In the case of $W > W_{c1}$, the DOS no longer vanishes at the Fermi level thus, gives rise to a DM phase \cite{k_dependent_SCBA, Mikito_k_dependent_SCBA}. The critical disorder strength $W_{c1}$ and the corresponding phase boundary predicted by SCBA is shown as the filled blue circles and blue line in Fig. \ref{phase_diagram}(b), respectively. This phase boundary separating the WSM1 phase and the DM phase is not quantitatively consistent to the results of the transfer matrix method (Such quantitative inconsistency for SCBA also has been reported for WSM1 without tilt \cite{SCBA_factor_2}). Besides, the SCBA cannot provide the intermediate 3D QAH phase.

However, in case of $b_t>1$, the DOS at the Fermi level is non-vanishing even in the clean limit. Therefore, one can no longer find such critical disorder strength $W_{c1}$ (see Fig. \ref{SCBA}(c)). It indicates that the WSM2 phase and the DM phase cannot be distinguished through SCBA.

Let us now move on to the real part of the self-energy. Since $Re{\{\Sigma_0\}}$ is $k$-independent, it simply shifts the Fermi level other than renormalizes the tilt term $a_t$ and $b_t$. (For Gaussian disorder potential, phase transition from WSM1 to WSM2 due to the renormailzation of tilt strength takes place after disorder-induced phase transition from WSM1 to DM phase \cite{k_dependent_SCBA}.) In the case of $b_t=0$, $\Sigma_0$ vanishes because of the presence of antiunitary particle-hole symmetry. For non-zero $b_t$, the disorder potential is not equivalently distributed above and below the Fermi level, hence the Fermi level is shifted by disorder effect (see Fig. \ref{SCBA}(d)). Although the numerical results show that the Fermi level $E_F$ is not shifted by $Re{\{\Sigma_0\}}$ significantly, however, this Fermi level shift may lead to the system into a Weyl metal other than a semimetal. Thus, the non-zero DOS at the Fermi level can be partially explained by the Fermi level shift.


The mass renormalization $Re{\{\Sigma_z\}}$ is negative and decreases with increasing of the disorder strength $W$. Besides, $Re{\{\Sigma_z\}}$ does not depend on the tilt term $b_t$ significantly so that it can be fitted by function $Re{\{\Sigma_z\}} = -0.02 W^2$ (see Fig. \ref{SCBA}(e)). Hence the critical disorder strength which separates the 3D QAH phase and the WSM phase is $W_{c2} = \sqrt{(1+m_z)/0.02} \approx 8.4$. According to the fitting function, the phase diagram predicted by SCBA is shown as Fig. \ref{phase_diagram}(b). The inconsistency between Fig. \ref{phase_diagram}(a) and \ref{phase_diagram}(b) can be explained by the fact that SCBA only works well in the small-$W$ region, therefore the SCBA-predicted 3D QAH phase in the region of $W>W_{c2}$ is absent in the results of the transfer matrix method. Besides, the NI phase is induced by the Anderson localization, which also cannot be described by the scheme of SCBA.

It is known that quantum lifetime and transport lifetime are two characteristic time scales \cite{Lifetime_ratio_Das_Sarma, DiracSM_lifetime_ratio, Qing-Dong} for metallic systems. Normally, these two time scales are of the same order for most 3D materials \cite{Lifetime_ratio_Das_Sarma}. However, experiments have shown that the ratio between transport lifetime and quantum lifetime is extremely large ($\sim 10^2$) for WSM1 \cite{WSM1_lifetime_ratio} when Fermi level approaches the Weyl nodes. Noticing that WSM2 possesses longer localization length (along the direction of tilt) and shorter quantum lifetime (see Fig. \ref{SCBA}(a), (b)) than WSM1, which indicates that the lifetime ratio (along the direction of tilt) for WSM2 is even larger (e.g. For disorder strength $W=8$, lifetime ratio at $b_t=2$ is $\sim 10^1$ times larger than $b_t=0$, see Fig. \ref{TMM_result}(a), (f) and Fig. \ref{SCBA}(a), (b)). Quantum lifetime and transport lifetime can be measured by Shubnikov$\textrm{-}$de Haas (SdH) oscillations and the electron mobilities, respectively \cite{DiracSM_lifetime_ratio}. This enormous lifetime ratio is expected to be observed for WSM2 as an exotic transport property.


\section{Summary and discussions}

In summary, we investigated the disorder-induced multiple phase transitions for disordered WSM in a wide range of tilt strength $0 \leq b_t \leq 2$. For disordered WSM1, an intermediate 3D QAH phase is confirmed between the WSM1 phae and DM phase. Furthermore, this 3D QAH phase is better seen with a non-zero $b_t$. For disordered WSM2, the DM phase is identified by the transfer matrix method, however, the intermediate 3D QAH phase is absent there. Though non-zero $b_t$ breaks the particle-hole symmetry, the phase diagram shows that the multiple phase transitions are clearly different in the region of $b_t<1$ and $b_t>1$. Besides, a longer localization length along the direction of tilt is observed for WSM2, leading us to believe that WSM2 is more robust against disorder than WSM1. Experimentally, an enormous lifetime ratio along the direction of tilt is expected to be a distinct transport property of WSM2. Finally, the critical disorder strength of the Anderson transition approaches to $0$ when $b_t$ is close to $1$. Therefore, remarkably, infinitely small disorder will induce an insulating phase near the boundary between the WSM1 and WSM2 phases.

The disorder-induced phase transition from the WSM2 phase to the DM phase cannot be described by the SCBA since both phases have non-zero DOS at the Fermi level. It remains an open question whether this DM phase can be confirmed by other results such as the Hall conductance \cite{Chui-Zhen_PRL, XiangRong_Wang} or the universal conductance fluctuations \cite{UCF}. Inspired by the insulating phase near $b_t=1$, another open question is that in real materials, whether the electron near the Weyl node will be localized by an infinitesimal disorder due to the scattering between the Weyl nodes and the possible nearly flat band.


\section*{ACKNOWLEDGMENTS}
We thank Qing-Feng Sun, Chui-Zhen Chen, Qing-Dong Jiang and Ziqiang Wang for helpful discussion. This work is financially supported by NBRPC (Grant No. 2015CB921102, 2014CB920901)
and NSFC (Grants Nos. 11534001, 11374219, 11504008) and  NSF of Jiangsu
Province, China (BK20160007).

\nocite{*}

\bibliography{disorder_WSM2}

\end{document}